\documentclass[aps, prstab, reprint, floatfix, showpacs]{revtex4-1}

\usepackage{graphicx}
\usepackage{booktabs}
\usepackage{notoccite}
\usepackage{threeparttable}
\newcommand{\unit}[1]{\ensuremath{\, \mathrm{#1}}}
\providecommand{\e}[1]{\ensuremath{\times 10^{#1}}}

\begin{document}

\title{Sensitivity of Niobium Superconducting Cavities to Trapped Magnetic Flux Dissipation}

\author{D.~Gonnella}
\email{dg433@cornell.edu}
\author{J.~Kaufman}
\author{M.~Liepe}
\email{mul2@cornell.edu}
\affiliation{Cornell Laboratory for Accelerator-Based Sciences and Education, Ithaca, NY, 14853, USA}


\date{\today}

\begin{abstract}
Future particle accelerators such as the the SLAC ``Linac Coherent Light Source-II'' (LCLS-II) and the proposed Cornell Energy Recovery Linac (ERL) require hundreds of superconducting radio-frequency (SRF) cavities operating in continuous wave (CW) mode. In order to achieve economic feasibility of projects such as these, the cavities must achieve a very high intrinsic quality factor ($Q_0$) to keep cryogenic losses within feasible limits. To reach these high $Q_0$'s in the case of LCLS-II, nitrogen-doping has been proposed as a cavity preparation technique. When dealing with $Q_0$'s greater than 1\e{10}, the effects of ambient magnetic field on $Q_0$ become significant. Here we show that the sensitivity to RF losses from trapped magnetic field in a cavity's walls is strongly dependent on the cavity preparation. Specifically, standard electropolished and 120$^\circ$C baked cavities show a residual resistance sensitivity to trapped magnetic flux of $\sim0.6$ and $\sim0.8$ n$\Omega$/mG trapped, respectively, while nitrogen-doped cavities show a sensitivity of $\sim 1$ to 5~n$\Omega$/mG trapped. We show that this difference in sensitivities is directly related to the mean free path of the RF surface layer of the niobium: shorter mean free paths lead to less residual resistance sensitivity to trapped magnetic flux in the dirty limit ($\ell<<\xi_0$) while longer mean free paths lead to lower sensitivity in the clean limit ($\ell>>\xi_0$). These experimental results are also shown to have good agreement with recent theoretical predictions for pinned vortex lines oscillating in RF fields.
\end{abstract}

\maketitle

\section{Introduction}
Superconducting Radio Frequency (SRF) cavities are the accelerating force in many of today's modern particle accelerators \cite{Hasan}. Because of the need to operate at sub-atmosphere liquid helium temperatures, the cryogenic load of the cavities must be minimized. By developing cavities with increasing intrinsic quality factors ($Q_0$), the SRF community has consistently pushed the boundaries of cryogenic efficiency (dissipated power is inversely proportional to $Q_0$) \cite{Hasan2}. Lowering of the mean free path of the surface layer of niobium has been shown to decrease dissipated power. This has been done for example with 120$^\circ$C baking\cite{GigiBaking} to eliminate the high field Q slope typically observed above 25 MV/m and to reduce the surface resistance by as much as 50\%. 

New machines such as the Linac-Coherent Light Source-II (LCLS-II) \cite{LCLS} to be constructed at SLAC have even more ambitious $Q_0$ requirements than have been previously proposed. LCLS-II requires an operating gradient of 16 MV/m with a $Q_0$ of 2.7\e{10} at 2.0 K\cite{LCLS}. In order to meet such an ambitious goal, nitrogen-doping of the SRF cavities has been proposed\cite{Anna}. Nitrogen-doping consists of treating niobium cavities in a UHV furnace at high temperatures (typically 600-900$^\circ$C) in a low atmosphere (a few mTorr) of nitrogen gas. It has been used to improve the Q in the medium field region (15-20 MV/m) due to a lowering of the mean free path of the material \cite{DanN2}. 

However, with very promising benefits, nitrogen-doping also carries its own technical difficulties that must be addressed before the technology can be applied to a full machine. Firstly, nitrogen-doped cavities typically quench at lower fields than un-doped cavities. This phenomenon will not be further discussed here. Secondly, nitrogen-doped cavities have been shown to have higher RF losses from trapped magnetic flux than un-doped cavities\cite{DanIPACFlux}. Here we will focus on the second point and how this sensitivity to losses from trapped magnetic flux depends on cavity preparation. Due to the low surface resistance of nitrogen-doped cavities, additional losses from trapped magnetic flux can result in significant increases to the dissipated power. When a cavity enters the Meissner state, some ambient DC magnetic field will be expelled while some will remain trapped in the material \cite{Tinkham}. This ``trapped flux'' can cause additional losses in RF fields, resulting in an increase in the temperature-independent residual resistance. However, the exact dynamics of a given cool down strongly impacts the amount of ambient magnetic field that is trapped in the cavity walls. Romanenko et. al. showed that fast cool down in nitrogen-doped cavities produced better $Q_0$ performance than slow cool down \cite{AnnaCoolDown}. Since then, work at Cornell and FNAL has shown that large spatial temperature gradients are important for maximal flux expulsion during cool down \cite{AlexMagField,DanHTC}.

Understanding the dependence of sensitivity of RF losses from trapped flux is crucial now as $Q_0$ performance and specifications keep increasing. Nearly all modern cavity preparation techniques use niobium that has an surface RF layer with a lower mean free path than the bulk niobium.  However the impact of new cavity preparations on the residual surface resistance due to trapped flux is not well understood. Previous measurements have shown a dependence of the residual resistance sensitivity to ambient magnetic field on the cavity preparation. Specifically, Weingarten et. al. showed that cavities constructed of niobium sputtered on copper (much smaller mean free path than high RRR bulk niobium) had less sensitivity to ambient field \cite{Weingarten_thinfilms} using a similar method to what we will discuss here. Additionally, they showed that residual resistance scales linearly with ambient magnetic field. Now with nitrogen-doping, we have the ability to finely tune the mean free path of the material in order to fully study this sensitivity's dependence on the full material parameter space. To achieve this goal, a program has been developed at Cornell to study the effects of ambient magnetic field on SRF cavities of differing preparation methods.

\section{Theoretical Considerations}

Vallet et. al. discuss the dependence of residual resistance on trapped ambient magnetic field in SRF cavities based on a phenomenological result \cite{Vallet}. It can be shown that the residual surface resistance, $R_0$, increases with trapped DC magnetic field. A trapped magnetic field, $B_{trapped}$ in an area $A$ breaks up into $N$ fluxoids each with a quantum of flux, $\phi_0$, so that
\begin{equation}
AB_{trapped}=N\phi_0.
\label{eq1}
\end{equation}
The size of these normal conducting cores is directly related to the coherence length, $\xi$. The additional residual resistance from the trapped magnetic flux can be estimated by the following simple model. We estimate the contribution to surface resistance from the normal conducting vortex cores. This goes by multiplying the number of fluxoids times the normal conducting resistance, $R_n$, times the fraction of the normal conducting area.
\begin{equation}
R_0 = N\frac{\pi\xi^2}{A}R_n=\frac{\pi\xi^2R_n}{\phi_0}B_{trapped}.
\label{eq2}
\end{equation}
Using the upper critical field, $B_{c2}$,
\begin{equation}
B_{c2}=\frac{\phi_0}{2\pi\xi^2},
\label{eq7}
\end{equation}
equation \ref{eq2} can be rewritten as
\begin{equation}
R_0=\frac{B_{trapped}}{2B_{c2}}R_n.
\label{eq3}
\end{equation}
This model predicts that as mean free path decreases (resulting in $B_{c2}$ increasing), sensitivity to trapped flux would decrease \cite{Hasan}. This is consistent with results on copper cavities coated with low RRR niobium \cite{Weingarten_thinfilms}. Specifically, for 1.3 GHz cavities with a $B_{c2}$ of 240 mT and a normal state resistance $R_n$ of 1.5 m$\Omega$ (typical for clean niobium \cite{Hasan}), this model suggests an additonal $\sim0.3$ n$\Omega$/mG of ambient magnetic field. This is in good agreement with experimental results on clean bulk niobium cavities in which a sensitivity of surface resistance to trapped magnetic flux was found to be 0.35 n$\Omega$/mG \cite{Vallet}.

Gurevich et. al. developed a more advanced theoretical model studying the impact of vortices on surface resistance for fields perpendicular to the surface \cite{Gurevich2013}. By modeling oscillations of vortex lines between pinning sites the additional surface resistance as a function of mean free path ($\ell$) can be computed beginning from London theory. The dissipated power from a single vortex line is
\begin{equation}
P=\frac{H_p^2\phi_0^2\left(\sinh\sqrt{2\nu}-\sin\sqrt{2\nu}\right)\sqrt{\nu}}{2^{3/2}\eta\ell_p\left(\cosh\sqrt{2\nu}+\cos\sqrt{2\nu}\right)},
\label{eq4}
\end{equation}
where $H_p$ is the RF peak magnetic field, $\phi_0$ is the flux quantum and $\ell_p$ is the mean spacing between pinning centers, with
\begin{equation}
\nu=\omega\eta\ell_p^2/\epsilon, \quad \epsilon=\frac{\phi_0^2g}{4\pi\mu_0\lambda^2}, 
\label{eq5}
\end{equation}
\begin{equation}
\eta=\phi_0B_{c2}/\rho_n, \quad g=\ln\kappa+\frac{1}{2}.
\label{eq6}
\end{equation}
The mean spacing must be much greater than the coherence length in order for the theory to be valid. Here $\lambda$ is the penetration depth, $\kappa$ is the Ginzburg-Landau parameter, $\rho_n$ is the normal conducting resistivity, $\omega$ is the RF frequency multiplied by $2\pi$, and $B_{c2}$ is the upper critical field of the superconductor defined by equation \ref{eq7}. $\rho_n$ is proportional to $1/\ell$ with a constant of proportionality for niobium of 0.37\e{-15} $\Omega\unit{m}^2$ \cite{goodman_rho} (Somewhat different values have been reported elsewhere \cite{RRR_ttc}.) The penetration depth \cite{pippard} and the coherence length can be calculated as a function of the mean free path \cite{Orlando}.

From the dissipated power per flux line (equation~\ref{eq4}), the total additional residual resistance from a DC magnetic field of magnitude $B_{trapped}$ can be computed by the relation
\begin{equation}
R_0=\frac{2PB_{trapped}}{\phi_0H_p^2},
\label{eq8}
\end{equation}
This prediction leads to two distinct regions with very different behavior. In the clean limit, the spacing between pinning centers, $\ell_p$, does not impact the surface resistance. In this region, losses decrease with longer mean free paths as $1/\sqrt{\ell}$. In the very dirty limit, the dependence is strongly impacted by $\ell_p$. If $\ell_p$ is a constant, surface resistance will increase very fast ($R_0\propto1/\ell^2$) with decreasing mean free path. It is reasonable to assume however that vortex lines will be pinned at defects. Therefore we propose a linear relationship between the mean spacing and the mean free path:
\begin{equation}
\ell_p=C\ell,
\label{eq9}
\end{equation}
with constant of proportionality, $C$. We will show that this assumption is in good agreement with experimental results. In this case the surface resistance will decrease with shorter mean free paths ($R_0\propto\ell$). The turning point between these two regions is heavily dependent on the constant of proportionality chosen. This behavior in the dirty limit is qualitatively similar between the two models discussed here and explains experimental results in which cavities constructed of niobium sputtered on copper (with much lower mean free paths and thus also lower mean spacing between pinning centers) showed less sensitivity to trapped ambient magnetic field than cavities made of bulk high RRR niobium (with higher mean free paths) \cite{Weingarten_thinfilms}.

\section{Experimental Results}

\subsection{Experimental Setup and Procedures}
Ten SRF cavity preparations were completed on six individual cavities to vary the mean free path of the RF penetration layer by impurity doping. These cavities were all made using the same fine grain RRR 320 material. Each preparation's sensitivity of residual resistance to ambient magnetic field was studied. All six cavities were 1.3 GHz ILC shaped single-cell cavities \cite{Tesla_cavity}. Six of the cavity preparations consisted of the same nitrogen-doping at 800$^\circ$C followed by a different amount of final vertical electropolishing (VEP) between 6 and 40 $\mu$m. Two of these five cavities had their surfaces reset and were prepared with nitrogen-doping at 900$^\circ$C and 990$^\circ$C, respectively, to increase their doping level. In the the cases of all doped cavities, the doping layer is very uniform over the length of the penetration depth \cite{DanLinacLCLS}. The ninth and tenth preparations were standard cavity preparations: VEP and VEP+120$^\circ$C bake. These exact details of all of these preparations are summarized in Table~\ref{tab1}.

\begin{table*}[tbh]
\centering
\begin{threeparttable}
\caption{Sensitivity of Cavities to Trapped Flux}
\begin{tabular}{|c|c|c|c|c|c|} \hline
\textbf{Cavity} & \textbf{Preparation} & \textbf{$T_c$ [K]} & \textbf{$\Delta/k_BT_c$} & \textbf{Mean Free} & \textbf{Sensitivity of $R_0$ to} \\
 & & & & \textbf{Path [nm]} & \textbf{Trapped Flux [n$\Omega$/mG]} \\
\hline \hline
LT1-3 & 990$^\circ$C ``Over-Doping''\tnote{1} & $9.1\pm0.1$ & $2.05\pm0.01$ & $4\pm1$ & $3.2\pm0.5$ \\
\hline
LT1-2 & 900$^\circ$C ``Over-Doping''\tnote{2} & $9.1\pm0.1$ & $2.00\pm0.01$ & $6\pm1$ & $4.7\pm0.6$ \\
\hline
LT1-2 & N-Doping\tnote{3} + 6 $\mu$m VEP& $9.3\pm0.1$ & $1.88\pm0.01$ & $19\pm6$ & $3.7\pm0.9$ \\ 
\hline
LT1-3 & N-Doping\tnote{3} + 12 $\mu$m VEP& $9.3\pm0.1$ & $1.91\pm0.01$ & $34\pm10$ & $3.1\pm0.5$ \\
\hline
LT1-1 & N-Doping\tnote{3} + 18 $\mu$m VEP& $9.3\pm0.1$ & $1.88\pm0.01$ & $39\pm12$ & $2.5\pm0.6$ \\
\hline
LT1-4 & N-Doping\tnote{3} + 24 $\mu$m VEP& $9.2\pm0.1$ & $1.89\pm0.01$ & $47\pm14$ & $2.2\pm0.2$ \\
\hline
LT1-5 & N-Doping\tnote{3} + 30 $\mu$m VEP& $9.2\pm0.1$ & $1.88\pm0.01$ & $60\pm18$ & $1.87\pm0.08$ \\
\hline
LT1-5 & N-Doping\tnote{3} + 40 $\mu$m VEP& $9.2\pm0.1$ & $1.94\pm0.01$ & $213\pm64$ & $1.06\pm0.02$ \\
\hline
NR1-3 & 100 $\mu$m VEP\tnote{4} & $9.2\pm0.1$ & $1.81\pm0.01$ & $800\pm100$ & $0.6\pm0.1$ \\
\hline
NR1-3 & 100 $\mu$m VEP\tnote{4} + 48 hour 120$^\circ$C Bake& $9.2\pm0.1$ & $1.96\pm0.01$ & $120\pm36$\tnote{5} & $0.88\pm0.07$ \\
\hline \end{tabular}
\begin{tablenotes}
\item[1] 100 $\mu$m VEP, 800$^\circ$C in vacuum for 3 hours, 990$^\circ$C in 30 mTorr of N$_2$ for 5 minutes, 5 $\mu$m VEP.
\item[2] 100 $\mu$m VEP, 800$^\circ$C in vacuum for 3 hours, 900$^\circ$C in 60 mTorr of N$_2$ for 20 minutes, 900$^\circ$C in vacuum for 30 minutes, light EP.
\item[3] 100 $\mu$m VEP, 800$^\circ$C in vacuum for 3 hours, 800$^\circ$C in 60 mTorr of N$_2$ for 20 minutes, 800$^\circ$C in vacuum for 30 minutes.
\item[4] After bulk VEP, cavities received an 800$^\circ$C heat treatment in vacuum for three hours.
\item[5] The 48 hour 120$^\circ$C bake has been shown to affect only the mean free path in a fraction of the RF penetration layer, especially at temperatures near the critical temperature \cite{Alex120C}. Because our method of mean free path extraction averages over this entire layer, the exact mean free path value is difficult to extract.
\end{tablenotes}
\label{tab1}
\end{threeparttable}
\end{table*}

\begin{figure}[t]
\centering
\includegraphics[scale=.05]{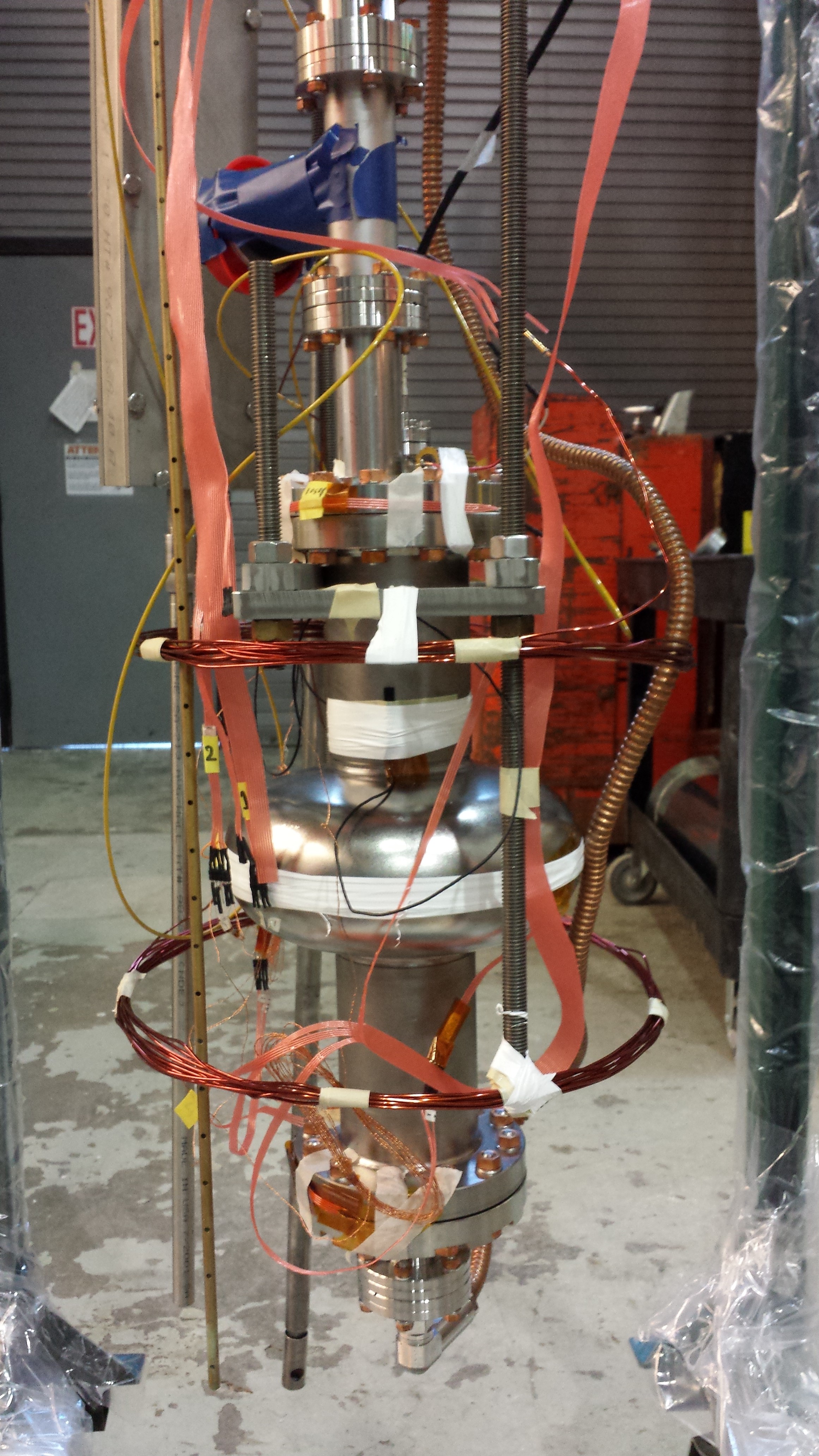}
\caption{The experimental setup. A 1.3 GHz ILC shaped single-cell cavity was surrounded in a Helmholtz coil to induce a uniform external magnetic field parallel to the cavity axis. A single-axis fluxgate magnetometor was placed on the cavity iris to measure magnetic field applied and trapped flux. Cernox temperature sensors were placed on the top and bottom flanges and on the equator to measure cool down rates and gradients.}
\label{fig1}
\end{figure}

\begin{figure}
\centering
\includegraphics[scale=.5]{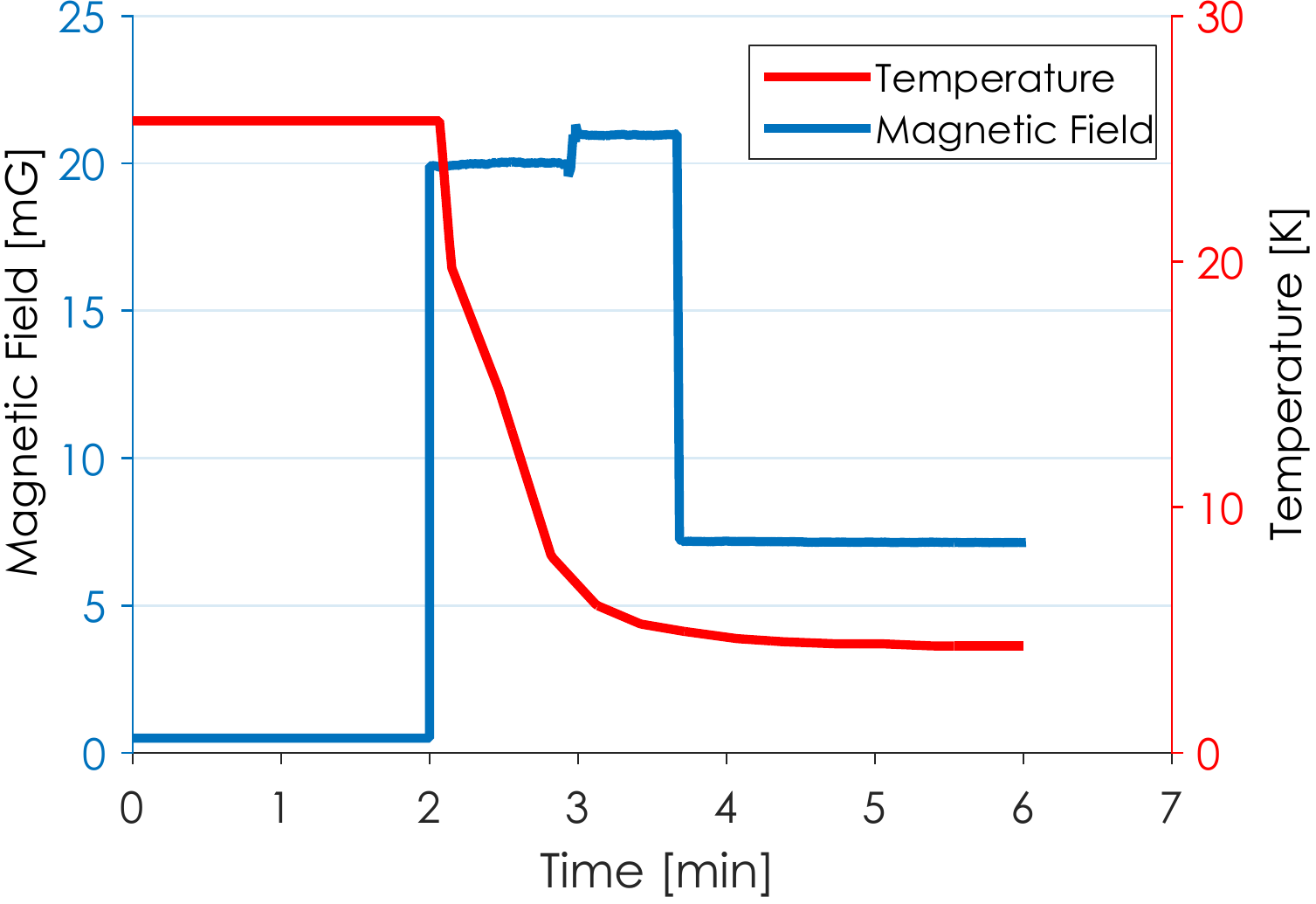}
\caption{A schematic of a typical cool down. The cavity sits above T$_c$ (at 25 K in this case) with the Helmholtz coil off. The external DC magnetic field is then turned on (to 20 mG in this case) and the cavity is cooled. There is a small jump in the measured magnetic field when the cavity becomes superconducting as flux is expelled. Once the cavity is cooled and superconducting, the magnetic field is turned off and it drops to a value which represents the amount of magnetic field trapped in the cavity walls (trapped flux).}
\label{fig2}
\end{figure}

For each cavity preparation described above, a variety of cool downs were conducted with different cool down speeds and applied DC external magnetic field. A Helmholtz coil was used to apply a uniform magnetic field parallel to the cavity axis. A fluxgate magnetometer located at the iris of the cavity and parallel to the applied magnetic field was used to measure both applied field and trapped magnetic flux. Three Cernox temperature sensors were used to measure temperature cooling rates and spatial gradients during the cool downs. A picture of the experimental setup is shown in Fig. \ref{fig1}. A typical cool down is shown in Fig. \ref{fig2}. First the cavity is above its critical temperature, T$_c$, with the coil off (in ambient magnetic field $B_{amb}$). Then the coil is turned on and the cavity is cooled. When the cavity transitions to the superconducting state, there is a small jump in the measured magnetic field as some flux is expelled. After cooling, the coil is turned off and the magnetic field drops to $B_{left}$, a value higher than the ambient field $B_{amb}$. The difference between these two fields is the amount of trapped flux:
\begin{equation}
B_{trapped}=B_{left}-B_{amb}.
\label{eq10}
\end{equation}
The trapped flux is the amount of magnetic field that is not expelled when the material becomes superconducting. The exact fraction of applied external magnetic field trapped depends strongly on the cool down mechanics: larger spatial temperature gradients giving less flux trapping\cite{dan_ipac_14_flux,dan_linac_14_flux}.

\subsection{Extraction of Material Properties}

\begin{figure}
\centering
\includegraphics[scale=.5]{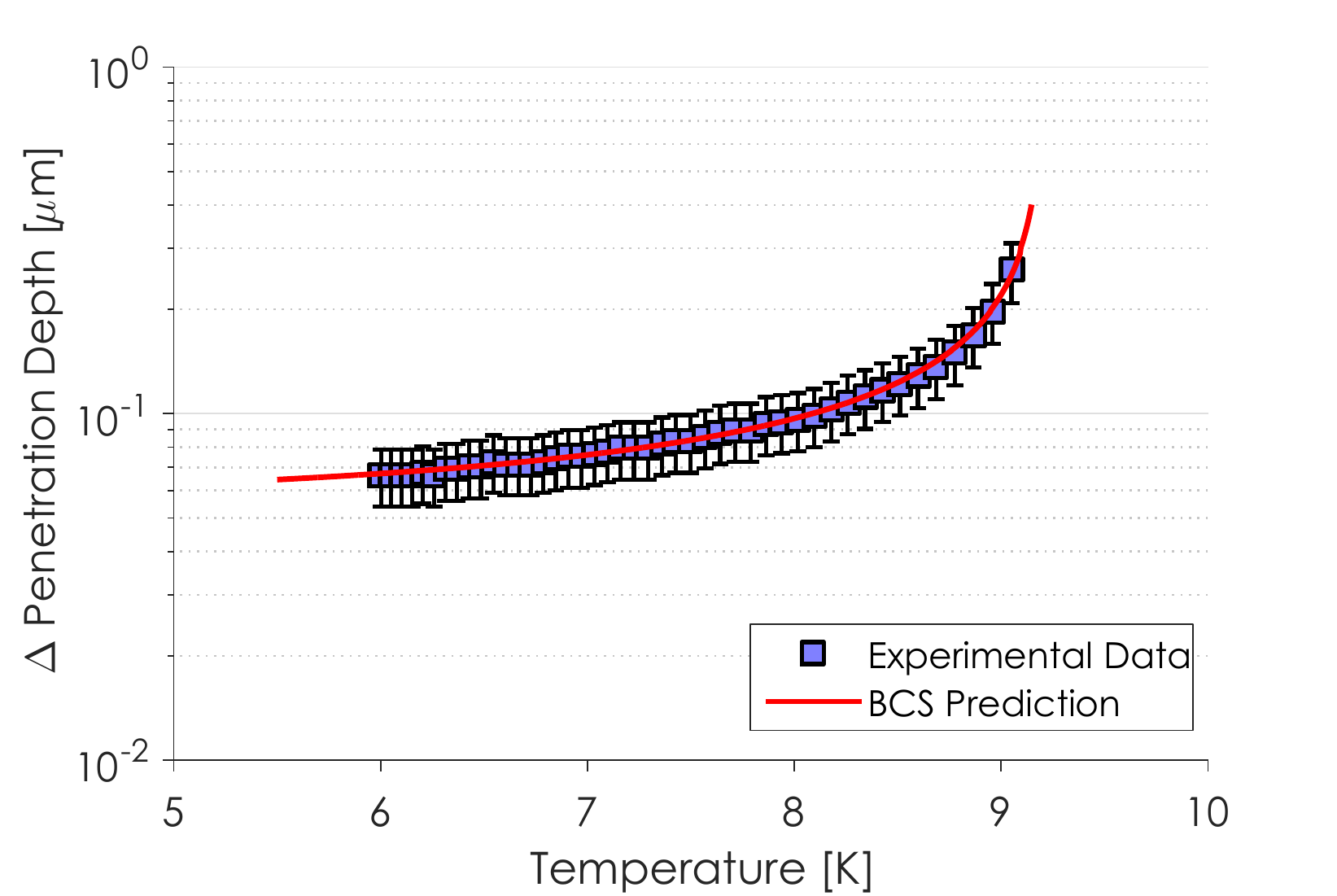}
\caption{An example of fitting of change in penetration depth vs temperature to extract mean free path.}
\label{fig5}
\end{figure}

\begin{figure}
\centering
\includegraphics[scale=.5]{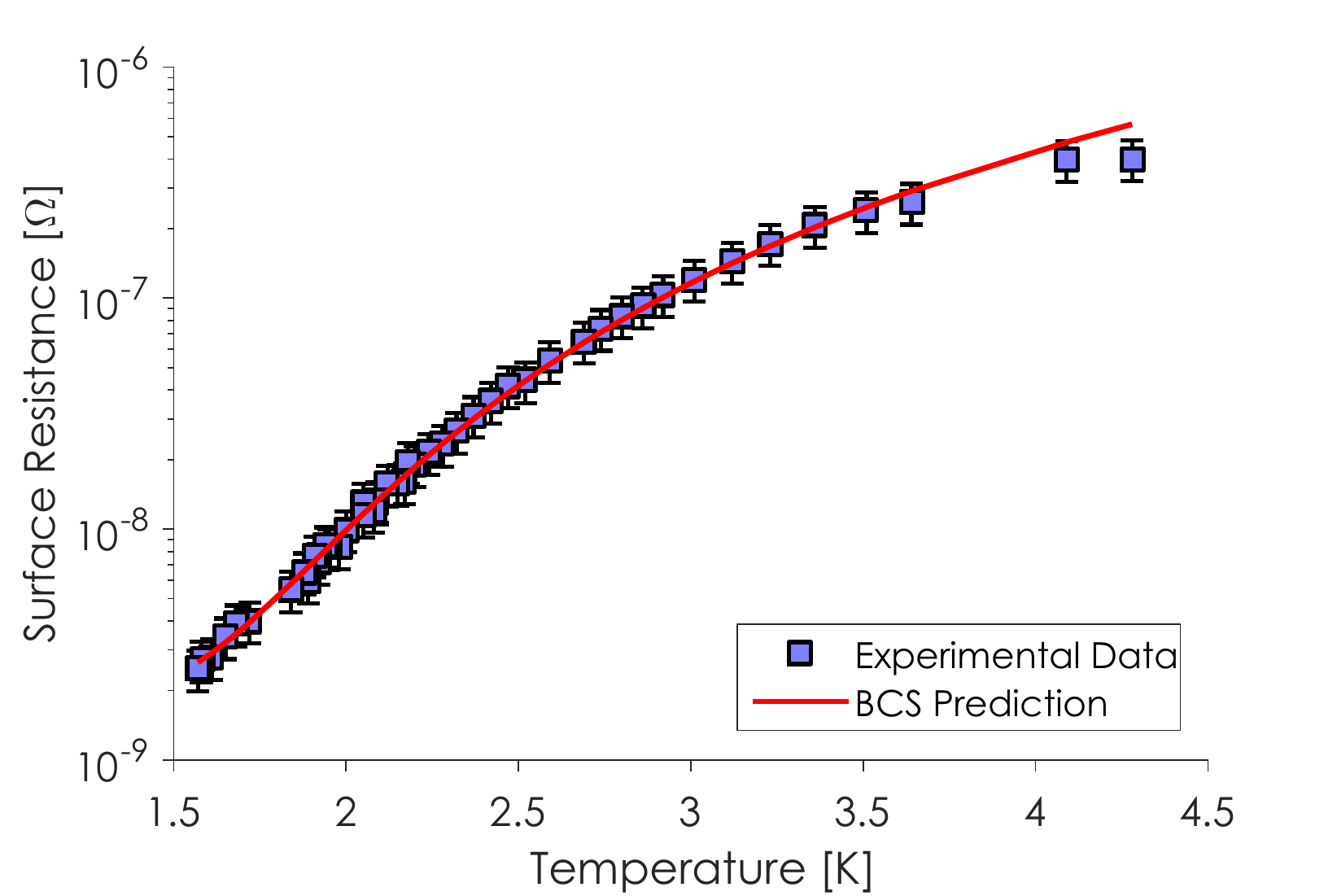}
\caption{An example of fitting surface resistance vs temperature to extract energy gap and residual resistance.}
\label{fig6}
\end{figure}

For each cool down, resonance frequency vs temperature was measured during warm up. Change in resonance frequency was converted to change in penetration depth using the method described in \cite{dan_srf_2013_highQ} and \cite{GigiBaking}. Additionally, $Q_0$ vs temperature was measured at low fields. From a combined fit of these two data sets using a polymorphic BCS analysis \cite{SRIMP}, residual resistance, energy gap, mean free path, and critical temperature were extracted. This provides an accurate method to extract these material properties since change in penetration depth vs temperature is sensitive to mean free path while surface resistance vs temperature is sensitive to energy gap. An example of this fitting is shown in Fig.~\ref{fig5} and Fig.~\ref{fig6}.

\begin{figure}
\centering
\includegraphics[scale=.5]{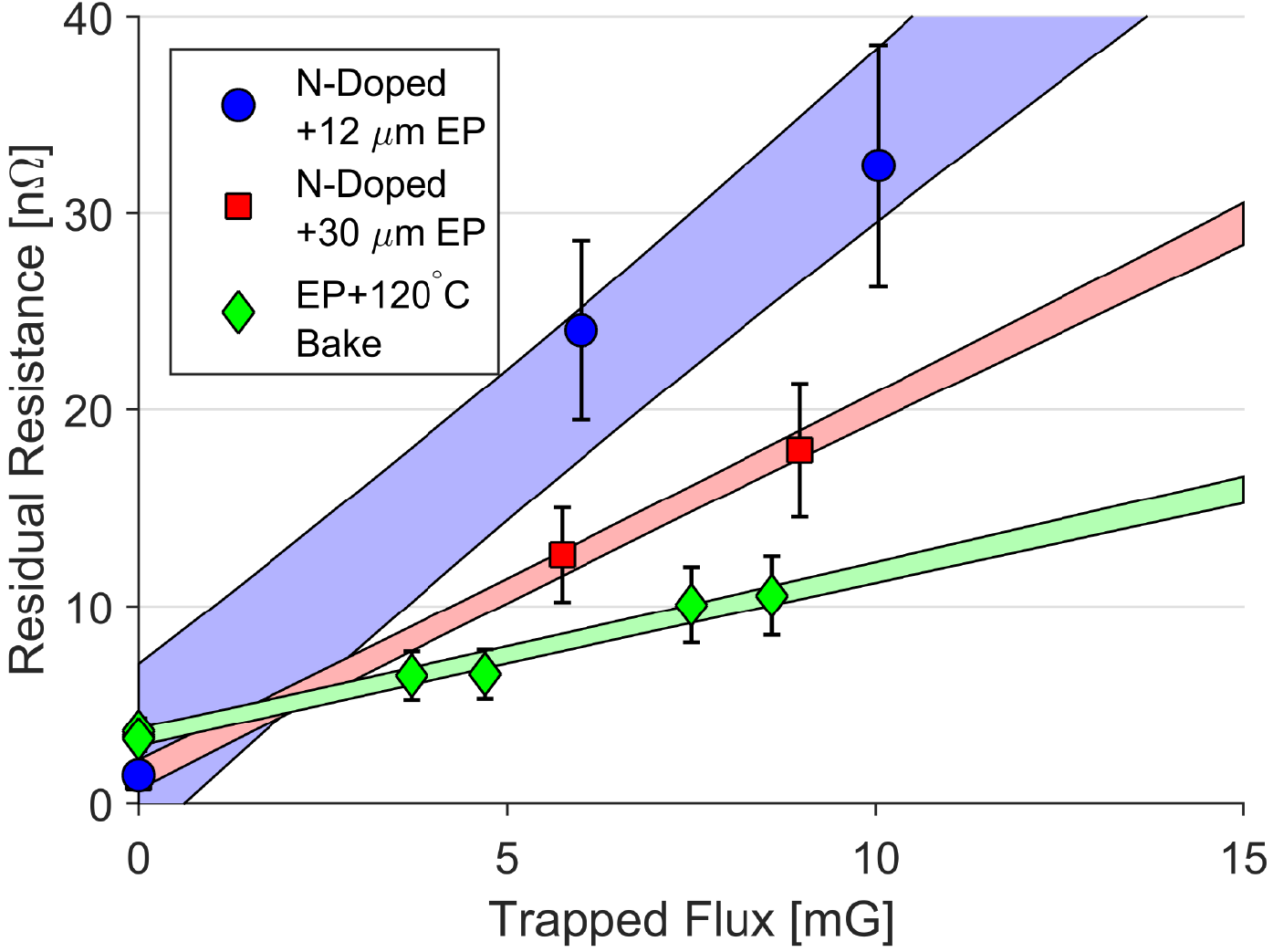}
\caption{The residual resistance as a function of trapped flux for two nitrogen-doped cavities and the EP+120$^\circ$C baked cavity. Nitrogen-doping results in a higher sensitivity to trapped flux than EP+120$^\circ$C baking. Also shown are linear fits with uncertainties in the fit.}
\label{fig3}
\end{figure}

For each cavity the residual resistance at low RF field ($<5$ MV/m) was found as a function of the trapped flux. Fig. \ref{fig3} shows results from three of the cavity preparations (two nitrogen-doped cavities and the VEP+120$^\circ$C baked cavity). We can see that residual resistance increases linearly with trapped flux as one would expect (see equations \ref{eq3} and \ref{eq8}): losses come directly from trapped vortices and doubling the trapped magnetic flux would double the trapped vortices. A linear fit was applied to the data and the slope extracted, giving the sensitivity of RF losses to trapped flux for each surface preparation. In addition to the extracted penetration layer material properties, these sensitivities are summarized in Table~\ref{tab1}.

We see that nitrogen-doping gives significantly smaller mean free paths than the VEP preparation. It is important to note that in the case of the VEP+120$^\circ$C bake cavity, the mean free path extracted will not be accurate. This is due to the 120$^\circ$C baking only affecting a small fraction of the RF penetration layer \citep{Alex120C}. Our method for extracting mean free path effectively averages over the whole penetration depth. This issue is not present in the VEP and nitrogen-doped cavities however since their surface layer is very uniform over several microns \cite{DanLinacLCLS}. The lowering of the mean free path in the doped preparations shows that the doping causes the niobium to become ``dirtier.'' This change is consistent with SIMS measurements on nitrogen-doped samples \cite{DanLinacLCLS}. We also see that for the same nitrogen-doping protocol (cavities doped at 800$^\circ$C), more material removal increases the mean free path of the material.

\subsection{Sensitivity of $R_0$ to Trapped Magnetic Flux}

From Fig.~\ref{fig3} and Table~\ref{tab1}, we can see that all eight nitrogen-doped surfaces had a higher sensitivity to trapped flux than the VEP and VEP+120$^\circ$C baked cavities. The sensitivity for the VEP preparation is similar to the previously predicted value of 0.3 n$\Omega$/mG as discussed in \cite{Vallet,Hasan}. For the same amount of trapped flux, a nitrogen-doped cavity will have a higher residual resistance and thus a lower $Q_0$. We also can see that there is a large spread in the sensitivities to trapped flux for the nitrogen-doped cavities. Stronger doping leads to larger sensitivities to trapped flux to a certain point, after which the sensitivity begins to decrease with even stronger doping.

The large difference in sensitivities for cavities of different preparations can be directly attributed to the mean free path of the materials. Looking at all extracted material properties in Table~\ref{tab1}, it is clear that there is no correlation between $T_c$ or $\Delta/k_bT_c$ with the sensitivity to trapped flux. There is however a clear correlation with the mean free paths of the preparations. This makes sense since doping effectively is changing the mean free path of the niobium by baking in impurities which act as scattering sites. 

\begin{figure}
\centering
\includegraphics[scale=.5]{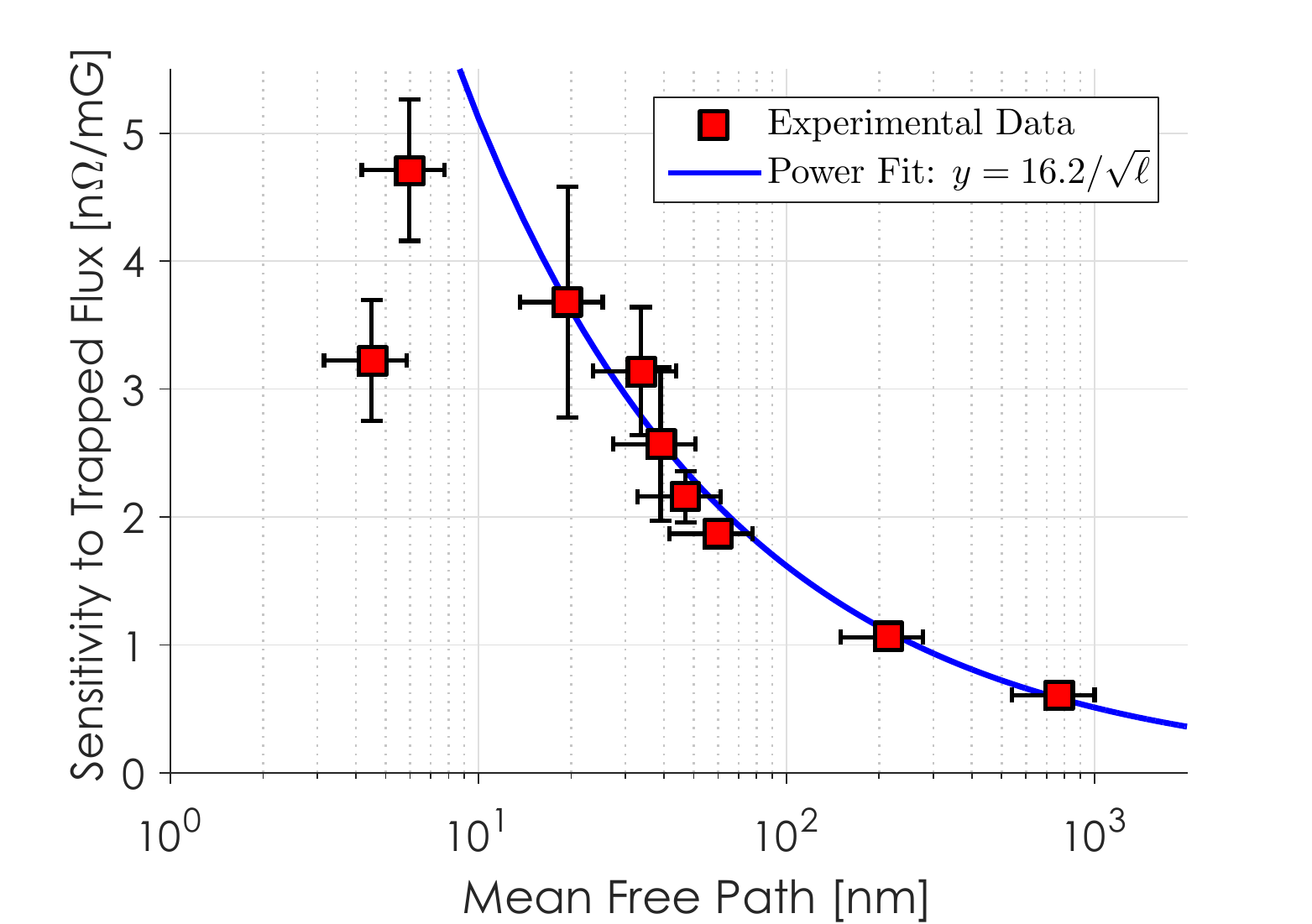}
\caption{Sensitivity of residual resistance to trapped flux as a function of mean free path for nine cavity preparations. Above mean free paths of 6 nm, longer mean free path resulted in lower sensitivity to trapped flux. Below 6 nm, sensitivity decreased with shorter mean free path. Also shown in a $1/\sqrt{\ell}$ fit as predicted by the Gurevich theory in the clean limit.}
\label{fig4}
\end{figure}

Figure \ref{fig4} shows the sensitivity of residual resistance to trapped flux (slope of Fig. \ref{fig3}) as a function of mean free path for nine of the ten cavity preparations. Due to the previously mentioned complications with the mean free path measurement for the VEP+120$^\circ$C bake preparation, this data point was omitted from Fig.~\ref{fig4}. We can see that for mean free paths above 6 nm, larger mean free path led to lower residual resistance sensitivity to trapped flux. Below mean free paths of 6 nm, the sensitivity decreased with smaller mean free path. Also shown is a least squares fit of $1/\sqrt{\ell}$ to the data with mean free paths above 20 nm as predicted by the Gurevich theory in the clean limit. We can see that for this region, the sensitivity does indeed change as $1/\sqrt{\ell}$.

\begin{figure}
\includegraphics[scale=.5]{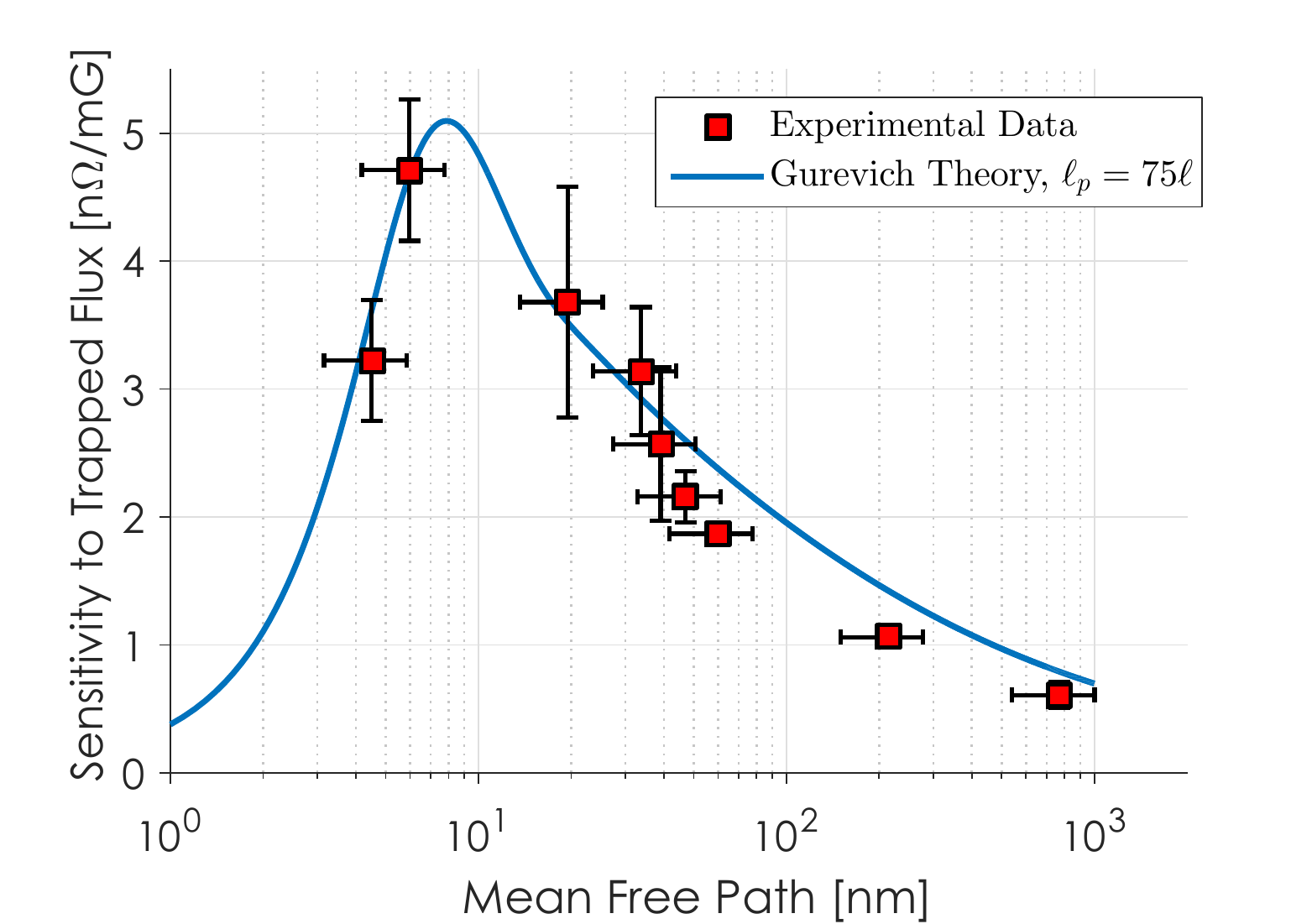}
\caption{Sensitivity of residual resistance to trapped flux as a function of mean free path. Shown in comparison with the experimental data is a theoretical prediction based on equation~\ref{eq8}. For this prediction, the mean spacing of pinning sites was assumed to be proportional to the mean free path. The constant of proportionality was found by fitting to the experimental data. For the relation between normal conducting resistivity and mean free path, $0.37\e{-15}$ was used \cite{goodman_rho}. Also used were the London penetration depth, $\lambda_L=39$ nm and the clean coherence length, $\xi_0=38$ nm \cite{Meservey,Poole}.}
\label{fig7}
\end{figure}

Referring back to equation~\ref{eq8}, we can compare the experimental results with the full theoretical prediction. Figure~\ref{fig7} shows the residual resistance sensitivity to trapped flux prediction from the Gurevich theory for a mean spacing of pinning sites $\ell_p=75\ell$. This value was chosen by fitting equation~\ref{eq8} to the experimental data varying the constant of proportionality, $C$, between the mean free path and the mean spacing between pinning sites. We can see from Fig.~\ref{fig7} that the theoretical model fits the data well in all regions thus supporting our assumption of a linear relationship between mean free path and mean spacing of pinning centers. These results show a maximum sensitivity at $\ell\approx8$ nm.It is likely that the constant of proportionality, $C$, depends on properties of the niobium used in cavity fabrication, for example, grain size.

\section{Conclusions}

Ambient magnetic fields during cool down can majorly impact the performance of SRF cavities. It has been well understood that magnetic fields can cause trapped flux that leads to additional residual resistance which in turn lowers the $Q_0$. We have shown that the exact residual resistance sensitivity to trapped flux is strongly dependent on the material preparation and especially the mean free path. As the surface gets ``dirtier,'' sensitivity will increase to a point after which increasing dirtiness leads to less sensitivity to trapped flux due to lowering of the mean spacing between pinning centers. These findings are consistent with the theoretical model based on losses from vortex line oscillations by Gurevich et. al.  

The findings shown here highlight concerns that must be considered in future machines in order to reach necessary high quality factors in cavities using new preparation methods such as nitrogen-doping. Since nitrogen-doped cavities have a smaller mean free path than more typically prepared cavities, their sensitivity to trapped flux can be much higher depending on how strongly doped they are. In theory, it is possible to dope strong enough that sensitivity to trapped flux will be on the order of EP and EP+120$^\circ$C bake cavities, however this is likely to bring with it a severe lowering of the quench field \cite{ipac2015_dan_ndoped}. Alternatively, the smallest possible doping levels should be used to lower the sensitivity of residual resistance to trapped flux. In order to maintain the very significant benefits that nitrogen-doping brings to SRF cavities, steps should be taken in order to either reduce ambient magnetic fields in a cryomodule or cool down in a manner that minimizes flux trapping (such as cooling with large transverse spatial temperature gradients\cite{DanHTC}). By understanding the dynamics of flux trapping along with how trapped flux impacts a cavity's residual resistance we are posed to reach intrinsic quality factors that are higher than have been reached before.

\section{Acknowledgments}
The authors would like to thank Newman Lab technical staff, especially Terri Gruber, Holly Conklin, and Greg Kulina for assistance with cavity preparation and cryogenic operations. We would also like to thank Alex Gurevich for helpful discussions on theoretical work relating to our experiment. Finally we would like to thank Anna Grassellino and Martina Martinello for discussions regarding losses at very small mean free paths. Work supported by the US DOE LCLS-II High Q Project and NSF grant PHY-1416318.

\bibliographystyle{unsrt}
\bibliography{bibtex_library}

\end{document}